\documentclass[conference, doublecolumn, hidelinks, xcolor=dvipsnames]{IEEEtran}

\pdfoutput=1
\IEEEoverridecommandlockouts

\usepackage{acronym}
\usepackage{amsmath}
\usepackage{amssymb}
\usepackage[english]{babel}
\usepackage{caption}
\usepackage{cite}
\usepackage[shortlabels,inline]{enumitem}
\usepackage{etoolbox}
\usepackage{graphicx}
\usepackage{hyperref}
\usepackage{bookmark}
\usepackage{ifthen}
\usepackage{cleveref}
\usepackage[utf8]{inputenc}
\usepackage{lipsum}
\usepackage{siunitx}
\usepackage[dvipsnames]{xcolor}
\usepackage[T1]{fontenc}
\usepackage[subnum]{cases}

\usepackage{epstopdf}
\graphicspath{{./fig/}}
\usepackage[toc,symbols]{glossaries}
\usepackage{bm}

\newtheorem{prop}{Proposition}

\newcommand{\Figure}[1]{Fig.~\ref{#1}}

\newcommand{\Equation}[1]{\eqref{#1}}
\newcommand{\Equations}[2]{\eqref{#1} and~\eqref{#2}}

\newcommand{\Table}[1]{Table~\ref{#1}}

\newcommand{\Section}[1]{Section~\ref{#1}}

\newcommand{\dx}{\ensuremath{\mathrm{d}x}}

\newcommand{\dzr}{\ensuremath{\mathrm{d}z_{\mathrm{RX}}}}
\newcommand{\dzm}{\ensuremath{\mathrm{d}z_{\mathrm{TX}}}}

\newcommand\given[1][]{\:#1\vert\:}
\newcommand\prob[1]{\textnormal{Pr}\{{#1}\}}
\newcommand\fpdf[2][]{\textnormal{f}_{\mathrm{#1}}\left(#2\right)}

\newcommand{\binomial}[1]{{\ensuremath{\textnormal{Binom}\Big({#1}\Big)}}}

\newcommand{\RXsimp}{$\mathrm{RX}_{\mathrm{simp}}$}

\renewcommand{\arraystretch}{1}
\acrodef{1D}[1-D]{one-dimensional}
\acrodef{2D}[2-D]{two-dimensional}
\acrodef{3D}[3-D]{three-dimensional}
\acrodef{MC}{molecular communication}
\acrodef{OOK}{ON-OFF keying}
\acrodef{ISI}{inter-symbol interference}
\acrodef{IUI}{inter-user interference}
\acrodef{IR}{impulse response}
\acrodef{ML}{maximum likelihood}
\acrodef{wlog}[w.l.o.g.]{without loss of generality}
\acrodef{BER}{bit error rate}
\acrodef{SER}{symbol error rate}
\acrodef{BSC}{Binary Symmetric Channel}
\acrodef{CDF}{cumulated density function}
\acrodef{UCA}{uniform concentration assumption}
\acrodef{AWGN}{Additive White Gaussian Noise}
\acrodef{PBS}{particle-based simulation}
\acrodef{MLSE}{maximum likelihood sequence estimator}
\acrodef{VE}{viterbi equalizer}
\acrodef{MLE}{maximum likelihood estimator}
\acrodef{CIR}{channel impulse response}
\acrodef{CIRs}{channel impulse responses}
\acrodef{SNR}{signal-to-noise ratio}
\acrodef{SDMA}{Space Division Multiple Access}
\acrodef{MCDMA}{Molecular Code Division Multiple Access}
\acrodef{ADMA}{Amplitude-Division Multiple Access}
\acrodef{MTDMA}{Molecular Time Division Multiple Access}
\acrodef{MDMA}{Molecular Division Multiple Access}
\acrodef{MIMO}{multiple-input multiple-output}
\acrodef{TDMA}{Time Division Multiple Access}
\acrodef{CDMA}{Code Division Multiple Access}
\acrodef{CSI}{channel state information}
\acrodef{TX}{transmitter}
\acrodef{TXs}{transmitters}
\acrodef{RX}{receiver}
\acrodef{RXs}{receivers}
\acrodef{ARE}{area rate efficiency}
\acrodef{w.r.t.}{with respect to}
\acrodef{GFPD}{green fluorescent protein Dreiklang}
\acrodef{EX}{eraser}

\DeclareMathOperator\erf{erf}

\makeatletter
\long\def\@makecaption#1#2{\ifx\@captype\@IEEEtablestring%
    \footnotesize\begin{center}{\normalfont\footnotesize #1}\\
        {\normalfont\footnotesize\scshape #2}\end{center}%
    \@IEEEtablecaptionsepspace
    \else
    \@IEEEfigurecaptionsepspace
    \setbox\@tempboxa\hbox{\normalfont\footnotesize {#1.}~~ #2}%
    \ifdim \wd\@tempboxa >\hsize%
    \setbox\@tempboxa\hbox{\normalfont\footnotesize {#1.}~~ }%
    \parbox[t]{\hsize}{\normalfont\footnotesize \noindent\unhbox\@tempboxa#2}%
    \else
    \hbox to\hsize{\normalfont\footnotesize\hfil\box\@tempboxa\hfil}\fi\fi}
\makeatother

\newcommand{\scaleSection}{\vspace*{-0.26cm}}
\newcommand{\scaleSubsection}{\vspace*{-0.24cm}}
\newcommand{\scaleSubsubsection}{\vspace*{-0.00cm}}

\newcommand{\scaleSubsectionBelow}{\vspace*{-0.08cm}}
\newcommand{\scaleSubsubsectionBelow}{\vspace*{-0.02cm}}
\newcommand{\scaleAlign}{\vspace*{-0.16cm}}

\begin{document}
\bstctlcite{disable_url}
\title{Media Modulation in Molecular Communications\vspace*{-0.4cm}
}
\author{\vspace*{-0.05cm}
\IEEEauthorblockN{Lukas Brand, Moritz Garkisch, Sebastian Lotter, Maximilian Schäfer, Kathrin Castiglione, and Robert Schober}
\IEEEauthorblockA{Friedrich-Alexander University Erlangen-Nuremberg, Germany}\vspace*{-0.8cm}
}

\maketitle
\begin{abstract}
In conventional molecular communication (MC) systems, the signaling molecules used for information transmission are stored, released, and then replenished by a transmitter (TX). However, the replenishment of signaling molecules at the TX is challenging in practice. Furthermore, in most envisioned MC applications, e.g., in the medical field, it is not desirable to insert the TX into the MC system, as this might impair natural biological processes. In this paper, we propose the concept of media modulation based MC where the TX is placed outside the channel and utilizes signaling molecules already existing inside the system. We consider signaling molecules that can be in different states which can be switched by external stimuli. Hence, in media modulation based MC, for information transmission, the TX stimulates the signaling molecules to encode information into their state. In particular, we elaborate media modulation for the group of photochromic molecules, which undergo light-induced reversible transformations, and study the usage of these molecules for information transmission in a three-dimensional duct system. We develop a statistical model for the received signal which depends on the distribution of the signaling molecules in the system, the reliability of the state control mechanism, and the randomness of molecule propagation. Furthermore, we analyze the performance of media modulation based MC in terms of the bit error rate (BER). We show that media modulation enables reliable information transmission, which renders a TX inside the channel unnecessary.

\end{abstract}

\setlength{\belowdisplayskip}{2pt}
\setlength{\belowdisplayshortskip}{2pt}

\acresetall
\scaleSection
\vspace*{0.2cm}
\section{Introduction}
\vspace*{-0.15cm}
To facilitate synthetic \ac{MC}, several concepts and modulation schemes for embedding information into molecular signals have been proposed over the last few years \cite{Jamali2019ChannelMF}.
In \cite{Kuran2011}, concentration shift keying (CSK) and molecular shift keying (MoSK), where information is encoded into the molecule concentration and the molecule type, respectively, have been proposed.
Moreover, the authors in \cite{Tang2021} proposed molecular type permutation shift keying (MTPSK) where information is encoded into permutations of different molecule types.
\par
For the implementation of these modulation schemes, the signaling molecules are stored, released, and then replenished by the transmitter (TX)\acused{TX}.
However, from a practical point of view, the replenishment of signaling molecules at the TX is difficult to realize, especially at microscale and in medical applications.
Therefore, it is desirable to develop alternative concepts that decouple the modulation process from the release of signaling molecules. \par
In this context, the concept of \textit{media modulation} has already been proposed for conventional wireless communication systems.
Here, information is embedded into the properties of the communication medium, i.e., the carrier signal is not directly modulated \cite{Khandani2013, Basar2019}.
Extending the concept of media modulation to MC, the authors of \cite{gohari2016information} propose to alter the properties of the channel to embed information. In particular, they advocate the use of chemotaxis or changing the flow velocity of the medium for modulation \cite{farahnak2020molecular}. In contrast to this channel-based form of media modulation, in this paper, we propose a new form of media modulation, where the properties of signaling molecules already present in the channel are modulated for information transmission via an external stimulus.

\vspace*{-1.2mm}In the simplest case, the molecules can be in two different states which can be switched by an external stimulus to encode the information to be conveyed. Such molecules may be naturally present in the environment or they may be injected and remain in the channel. Media modulation based MC can overcome several shortcomings of existing MC systems by
(i) avoiding repeated injections of signaling molecules, making the replenishment of the TX unnecessary, (ii) reducing the soiling of the channel due to the deposition of signaling molecules after repeated injection.
Furthermore, the TX, i.e., the source of the external stimulus, is not placed inside the channel and does not influence molecule propagation.
The practical feasibility of media modulation hinges on the availability of suitable switchable signaling molecules, of course.
The general concept of media modulation based \ac{MC} is new and has not been reported in the literature, yet.
Nevertheless, redox-based MC \cite{Kim2019} can be interpreted as an instance of media modulation where signaling particles are switched between two states, the reduced state and the oxidized state.
However, the focus of \cite{Kim2019} was on the biological and experimental aspects of redox-based MC, while a thorough communication theoretical analysis was not conducted.
Furthermore, in biological systems, a natural form of media modulation can be observed during phosphorylation, where a phosphoryl group is added to a protein affecting the properties of the protein. The phosphorylation is mediated by a kinase, a specific type of enzyme, which in turn can be controlled by an external stimulus \cite{grusch2014spatio, chang2014light}.

\vspace*{-1.2mm}Another promising candidate for signaling molecules for media modulation are photochromic molecules. These molecules can be reversibly interconverted between two states, where the transition between the states is induced by light \cite{balzani2014photochemistry}. Photochromic molecules are well established in molecular devices for information processing\cite[Chap.~10]{balzani2014photochemistry}, but their exploitation for media modulation in synthetic MC systems is new.
Photochromic systems facilitate different functionalities including writing, reading, and erasing of information by an external light stimulus.
In this paper, we develop a novel modulation scheme for synthetic MC systems employing photochromic signaling molecules.
Hereby, the information is embedded into the state of a photochromic molecule.
As photochromic signaling molecule, we exemplarily consider the reversibly photo-switchable green fluorescent protein variant ``Dreiklang'' (GFPD)\acused{GFPD}, whose fluorescence can be reversibly switched by light stimuli of mutually different wavelengths \cite{brakemann2011reversibly}. Fluorescence denotes the ability of a molecule to first absorb light and then radiate light back at a higher wavelength, i.e., with a lower energy, which makes it possible to read out the current state of a \ac{GFPD}.
In the proposed system, fluorescent and non-fluorescent GFPD correspond to state A and state B of the signaling molecule, respectively.
The fluorescence of GFPD can be switched on (B$\to$A) and off (A$\to$B) by light stimuli at different wavelengths.
Consequently, assuming the GFPD molecules are suspended in a liquid medium as elements of the envisioned synthetic MC system, optical sources emitting light at mutually different wavelengths can be used as TX (B$\to$A) and eraser (EX, A$\to$B)\acused{EX} units, respectively, and an optical sensor can be employed as receiver (RX)\acused{RX}.

As the proposed form of media modulation is studied for the first time in this paper, we focus on the communication theoretical modeling of the TX, while an in-depth study of the RX and the \ac{EX} is left for future work.
The main contributions of this paper can be summarized as follows:
\begin{itemize}
	\item We propose a new form of media modulation for synthetic MC and employ switchable photochromic molecules for information transmission. This novel concept does not require a \ac{TX} that stores and releases molecules, and therefore overcomes several drawbacks of existing modulation schemes, e.g., the need for TX replenishment.
	\item Since for media modulation, the state of photochromic molecules is switched by external light stimuli, we provide an in-depth investigation of the subjacent photochemical process. The analysis is exemplarily done for \ac{GFPD}.
	\item We derive an analytical channel model for the proposed system including the channel impulse response and a statistical model for the received signal. Moreover, we analyze the \ac{BER} of the resulting \ac{MC} system.
\end{itemize}
The remainder of this paper is organized as follows.
In Section~\ref{sec:model}, we introduce the considered MC system and describe the proposed media modulation scheme.
In Section~\ref{sec:math}, we derive an analytical end-to-end model for the proposed \ac{MC} system, and an expression for the \ac{BER} is provided in Section~\ref{sec:performance}.
In Section~\ref{sec:evaluation}, we evaluate the proposed models numerically.
Section~\ref{sec:conclusion} concludes the paper and outlines topics for future work.

\scaleSection
\vspace*{+0.05cm}
\section{System Model}
\label{sec:model}
In this section, we describe the proposed media modulation-based \ac{MC} system, including the modulation, propagation, and reception mechanisms. The system model presented in this section is generic and applicable to different types of photoswitchable fluorescent molecules. Later on, in \Section{sec:evaluation}, the model is specialized to \ac{GFPD}.
\begin{figure}[!tbp]
  \includegraphics[width=\columnwidth]{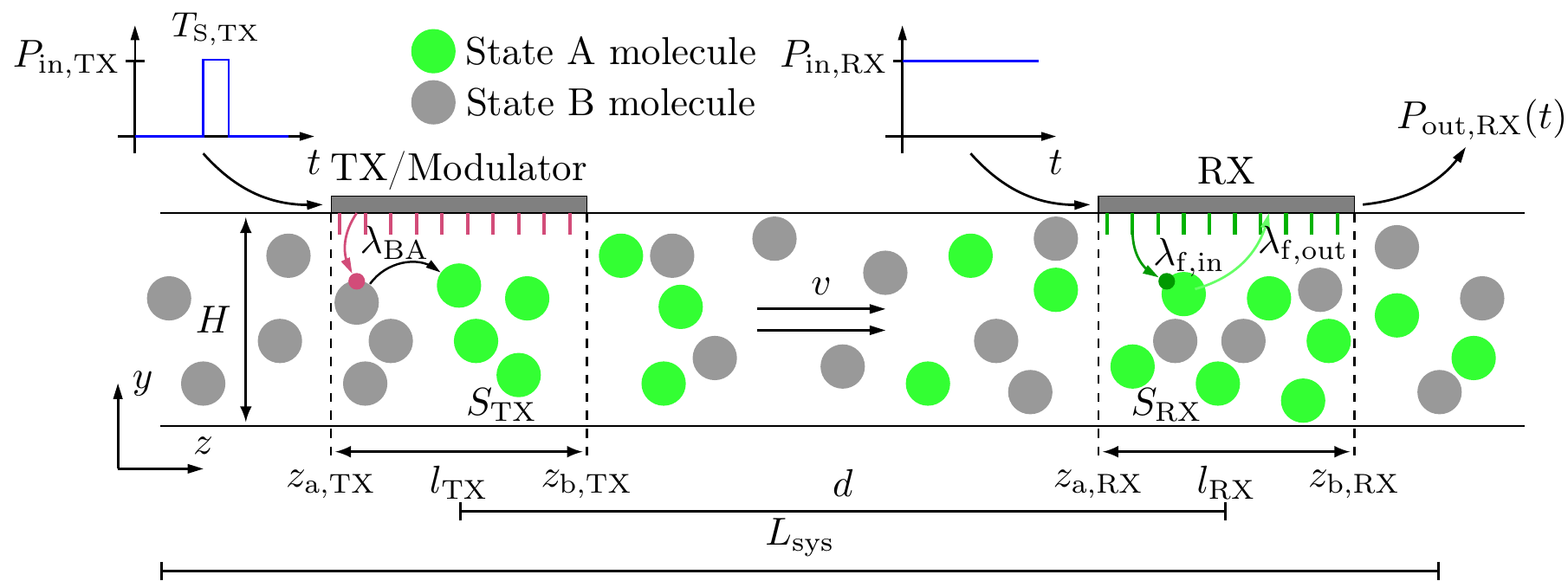}\vspace*{-0.05mm}
  \caption{\ac{MC} system model: Signaling molecules in state B (gray) are uniformly distributed in a \ac{3D} rectangular pipe of height $H$ and width $W$. For information transmission, the molecules are switched via a photochemical process ($\lambda_{\mathrm{BA}}$) to state A molecules (green) at the \ac{TX}, which can be controlled by an optical signal with input power $P_{\mathrm{in, TX}}$. The \ac{RX} triggers a fluorescence reaction in the state A molecules, which results in a measurable light power $P_{\mathrm{out, RX}}$. The molecules propagate by Brownian motion and uniform flow.}
  \label{system_model_picture}
  \vspace*{-0.5cm}
\end{figure}
\scaleSubsection
\vspace*{0.05cm}
\subsection{Topology, Modulation, Propagation, and Reception}\label{general_system_model}
\scaleSubsectionBelow
We consider a \ac{3D} straight rectangular duct with height $H$, width $W$, and infinite axial extent, which is oriented along the $z$-axis, cf. \Figure{system_model_picture}. The duct is filled with a fluid, which flows in $z$-direction with constant velocity $v>0$, i.e., we assume uniform flow, as is widely done in the \ac{MC} literature \cite{Jamali2019ChannelMF}. Moreover, we assume that molecules are reflected at the boundaries of the duct, i.e., the duct surface is impermeable to molecules. Additionally, at the duct sections where the \ac{TX} and the \ac{RX} are located, we assume the duct surface to be transparent to light. The communication process of interest takes place in a subvolume S of the duct of length $L_{\mathrm{sys}}$ and volume size $V_{\mathrm{sys}} = W H L_{\mathrm{sys}}$.

At time $t=0$, the fluid is well-mixed and $N_{\mathrm{sys}}$ signaling molecules are uniformly distributed in subvolume S. These signaling molecules are photochromic, i.e., they can assume two distinguishable states, A and B, and depending on their state, we will refer to them as state A molecules and state B molecules in the following, respectively. The state of a photochromic molecule can be changed by irradiation with light. We assume that the molecules are initially in state B, which might be the equilibrium state or can be enforced by a \ac{EX} upstream, which forces the molecules into state B.

\scaleSubsubsection
\subsubsection{Media modulation at the TX and EX}
\scaleSubsubsectionBelow
We consider a \ac{2D} \ac{TX} with area $A_{\mathrm{TX}} = l_{\mathrm{TX}} W$, which is attached to the outer boundary of the duct and extends in $z$-direction in the interval $[z_\mathrm{a,TX}, z_\mathrm{b,TX}]$, i.e., the \ac{TX} has an axial extent of $l_{\mathrm{TX}} = z_\mathrm{b,TX} - z_\mathrm{a,TX}$, cf. \Figure{system_model_picture}. For information transmission, at $t=0$, the \ac{TX} radiates light of wavelength $\lambda_{\mathrm{BA}}$ and power $P_{\mathrm{in, TX}}$ for an illumination duration of $T_{\mathrm{S,TX}}$ onto the volume $S_{\mathrm{TX}} = \{ (x,y,z) \in \mathrm{S} \given z_\mathrm{a,TX} \leq z \leq z_\mathrm{b,TX}\}$ of size $V_{\mathrm{TX}} = W H l_{\mathrm{TX}}$. The \ac{TX} uses \ac{OOK} modulation \cite{Jamali2019ChannelMF}, i.e., the \ac{TX} either radiates power $P_{\mathrm{in, TX}} = P_{\mathrm{in, TX, on}}$ or is inactive, i.e., $P_{\mathrm{in, TX}} = 0$, representing binary symbols $s = 1$ and $s = 0$, respectively. We assume that binary values $0$ and $1$ are equiprobable. The \ac{TX} radiation triggers a photochemical reaction that switches the state of a signaling molecule from B to A with probability $p_{\mathrm{switch}}$, see \Section{photoChemic}.
Since the focus of this paper is on the proposed novel modulation technique, we assume single symbol transmission, i.e., \ac{ISI} is not considered.

In the context of media modulation, the purpose of the \ac{EX} is to trigger the photochemical reaction that switches the state of a molecules from A back to B, i.e., the inverse reaction compared to the reaction at the \ac{TX}. Therefore, the \ac{EX} radiates light of a different wavelength $\lambda_{\mathrm{AB}}$, i.e., $\lambda_{\mathrm{AB}} \neq \lambda_{\mathrm{BA}}$. The geometry and functionalities of the \ac{EX} are similar to those of the \ac{TX}. However, details are omitted here, and the \ac{EX} is not explicitly modeled in this paper, cf. \Figure{system_model_picture}.

\scaleSubsubsection
\subsubsection{RX model}
\scaleSubsubsectionBelow
We consider a \ac{2D} \ac{RX} with area $A_{\mathrm{RX}} = l_{\mathrm{RX}} W$, which is attached to the outer boundary of the duct and extends in $z$-direction in the interval $[z_\mathrm{a,RX}, z_\mathrm{b,RX}]$, i.e., the \ac{RX} has an axial extent of $l_{\mathrm{RX}} = z_\mathrm{b,RX} - z_\mathrm{a,RX}$.  We assume that \ac{TX} and \ac{RX} have the same orientation and therefore the distance between the \ac{TX} and \ac{RX} centers is $d = \frac{z_\mathrm{b,RX}+z_\mathrm{a,RX}}{2} - \frac{z_\mathrm{b,TX}+z_\mathrm{a,TX}}{2}$. The \ac{RX} is capable of emitting light of wavelength $\lambda_{\mathrm{f, in}}$ and of sensing light of wavelength $\lambda_{\mathrm{f, out}}$. In particular, the \ac{RX} constantly radiates light of wavelength $\lambda_{\mathrm{f, in}}$ and power $P_{\mathrm{in, RX}}$ into the subvolume $S_{\mathrm{RX}} = \{ (x,y,z) \in \mathrm{S} \given z_\mathrm{a,RX} \leq z \leq z_\mathrm{b,RX}\}$ of size $V_{\mathrm{RX}} = W H l_{\mathrm{RX}}$. The radiated light triggers a fluorescence reaction at molecules that are in state A, i.e., the illuminated molecules in state A radiate light with wavelength $\lambda_{\mathrm{f, out}}$. This results in a received light power $P_{\mathrm{out, RX}}$ at the \ac{RX}. We note that all considered wavelengths, i.e., $\lambda_{\mathrm{BA}}$, $\lambda_{\mathrm{AB}}$, $\lambda_{\mathrm{f, in}}$, and $\lambda_{\mathrm{f, out}}$, are mutually distinct. For \ac{GFPD}, the values of these wavelengths are provided in \Table{Table:Parameter}.

As previously mentioned, we focus on the proposed novel modulation process at the \ac{TX}, and therefore employ a simple \ac{RX} model. In particular, we approximate the receiver by a transparent receiver \RXsimp. \RXsimp $\mkern+5mu$ is able to count the number of state A molecules in $S_{\mathrm{RX}}$ at fixed sampling time $t_{\mathrm{s}} = \frac{d}{v}$, i.e., the time for a molecule to propagate from \ac{TX} to \ac{RX} due to the uniform flow.

\scaleSubsubsection
\subsubsection{Propagation}
\scaleSubsubsectionBelow
In both states A and B, the molecules are subject to \ac{3D} Brownian motion characterized by diffusion coefficients $D_{\mathrm{A}}$ and $D_{\mathrm{B}}$, respectively, and uniform flow. While in general a chemical reaction can result in a change of the molecule structure and size, in the absence of experimental data, we assume $D_{\mathrm{A}} = D_{\mathrm{B}}$ for \ac{GFPD}, i.e., the size of the molecule remains unaltered when the state of the molecule changes. %Additionally, the molecules are transported along the $z$-axis by means of the uniform flow.\\
\scaleSubsection
\subsection{Photochemical Reaction}\label{photoChemic}
\scaleSubsectionBelow
We employ photochromic molecules as signaling molecules in our system, i.e., molecules that can be reversibly interconverted from state A to state B by external light. This allows reading, writing, and erasing of information embedded into the state of the molecule. In this paper, we focus on the modulation, i.e., the writing of information, by switching molecules in state B to state A.

The modulation process is described by the following chemical reaction \cite[Eqs.~(1.3),~(12.18),~(12.19)]{balzani2014photochemistry}
\scaleAlign
\vspace*{-0.1cm}
\begin{align}
  \mathrm{B } +  \varphi_{\mathrm{B}} P_{\lambda_{\mathrm{BA}}}\, \rightarrow \, \mathrm{A}\label{OFF_process},
\end{align}
where $P_{\lambda_{\mathrm{BA}}}$ denotes a photon with energy \cite[Eq.~(1.2)]{balzani2014photochemistry}
\scaleAlign
  \begin{align}
    E_{\mathrm{P_{\lambda_{\mathrm{BA}}}}} = h f_{\mathrm{BA}} = h \frac{c}{\lambda_{\mathrm{BA}}}\,.
    \label{eq:E_photon}
  \end{align}
  Here, $h$, $c$, and $f_{\mathrm{BA}}$ denote the Planck's constant, the speed of light, and the radiation frequency, respectively.
  The reaction quantum yield $\varphi_{\mathrm{B}}$ is defined as the ratio of the number of state B molecules, $N_{\mathrm{B}}$, in $S_{\mathrm{TX}}$ switched in unit time to the number of photons, $N_{\mathrm{P}}$, absorbed in unit time \cite[Eqs.~(12.18),~(12.20)]{balzani2014photochemistry}, i.e.,
  \scaleAlign
  \vspace*{-0.18cm}
\begin{align}
  \varphi_{\mathrm{B}} = -\frac{\mathrm{d}N_{\mathrm{B}}}{\mathrm{d}t} \Big/ \frac{\mathrm{d}N_{\mathrm{P}}}{\mathrm{d}t} \;.
    \label{quantumYield}
\end{align}
In the following, we assume that the molecules are approximately static during illumination, which is a valid assumption if the propagation distance of a molecule within the illumination duration, $T_{\mathrm{S,TX}}$, is much shorter than the \ac{TX} length $l_{\mathrm{TX}}$, i.e.,
\scaleAlign
  \vspace*{-0.25cm}
\begin{align}
  \underbrace{\sqrt{2 D_{\mathrm{B}} T_{\mathrm{S,TX}}}}_{\mathrm{Diffusion}} + \underbrace{v \, T_{\mathrm{S,TX}}}_{\mathrm{Flow}} \ll l_{\mathrm{TX}} \;,
  \label{eq:assumption_propagation_length}
\end{align}
where the propagation distance due to diffusion is characterized by its standard deviation $\sqrt{2 D_{\mathrm{B}} T_{\mathrm{S,TX}}}$. Inequality \Equation{eq:assumption_propagation_length} holds for typical system parameters, see \Table{Table:Parameter}.
Therefore, in the following, the reaction process is modeled for a closed volume, i.e., we assume that during modulation molecules do not move into or out of the volume.

Photons generated by light of wavelength $\lambda_{\mathrm{BA}}$ can only be absorbed by molecules in state B. Hence, the absorption of photons emitted by the \ac{TX} is governed by the Beer-Lambert law \cite[Eqs.~(12.21),~(12.22)]{balzani2014photochemistry}
\scaleAlign
\begin{align}
   \frac{\mathrm{d}N_{\mathrm{P}}(t)}{\mathrm{d}t} &= q_{\mathrm{in, TX}} \bigg(1-\exp\bigg(-\frac{ \log(10) \epsilon \, H N_{\mathrm{B}}(t)}{V_{\mathrm{TX}} N_{\mathrm{Av}}}\bigg)\bigg)\nonumber \\[-0.2cm]
   &\overset{{\mathrm{Eq.} \Equation{quantumYield}}}{=} - \frac{1}{\varphi_{\mathrm{B}}} \frac{\mathrm{d}N_{\mathrm{B}}(t)}{\mathrm{d}t} \;,
   \label{differential_equation}
\end{align}
where $q_{\mathrm{in, TX}}$, $\epsilon$, and $N_{\mathrm{Av}}$ denote the photon flux into $V_{\mathrm{TX}}$, the molar absorption coefficient in $\si{\meter^2 \mathrm{mol}^{-1}}$, and the Avogadro constant, respectively.
Note that the duct height, $H$, determines the maximum distance a photon can propagate through the volume.
The photon flux, $q_{\mathrm{in, TX}}$, is constant within $T_{\mathrm{S,TX}}$ and is given by
\scaleAlign
\vspace*{-0.3cm}
\begin{align}
  q_{\mathrm{in, TX}} = \frac{P_{\mathrm{in, TX}} A_{\mathrm{TX}} }{  E_{\mathrm{P_{\lambda_{\mathrm{BA}}}}}},
  \label{eq:Light_source_to_photons}
\end{align}
where all photons have the same energy $E_{\mathrm{P_{\lambda_{\mathrm{BA}}}}}$, as defined in \Equation{eq:E_photon}, due to the use of light with fixed wavelength $\lambda_{\mathrm{BA}}$.
We assume an initial number of molecules in state B of $N_{\mathrm{B}}(t=0) = N_{\mathrm{TX}}$ in $S_{\mathrm{TX}}$, cf. \Section{ssSec:InTX}. We note that $N_{\mathrm{TX}}$ is random, hence, $N_{\mathrm{TX}}$ can be different for every modulation process. Finally, the number of molecules in state B in $S_{\mathrm{TX}}$ as a function of time follows from \Equations{differential_equation}{eq:Light_source_to_photons} as follows
\scaleAlign
  \vspace*{-0.05cm}
\begin{align}
  N_{\mathrm{B}}(t)\mkern-4.5mu &= \mkern-4.5mu\frac{1}{a}\mkern-4.5mu \log\mkern-4.5mu\bigg[1\mkern-4.5mu-\mkern-4.5mu\exp\mkern-4mu\left(-\varphi_{\mathrm{B}} \,a \, q_{\mathrm{in, TX}}\, t \right)\mkern-4.5mu \Big(\mkern-4.5mu1-\mkern-4.5mu\exp\mkern-4mu\big( N_{\mathrm{TX}} a \big)\mkern-4.5mu\Big)\mkern-4.5mu\bigg] \mkern+2mu,\mkern-4mu\label{diff_solution}
\end{align}
where $a = \frac{\log(10) H \epsilon}{V_{\mathrm{TX}} N_{\mathrm{Av}}}$.
Finally, the probability of any molecule in $S_{\mathrm{TX}}$ to be switched from state B to state A within the irradiation time $T_{\mathrm{S,TX}}$ is given by
\scaleAlign
\vspace*{-0.1cm}
\begin{align}
  \hspace{-0.3cm}p_{\mathrm{switch}} = 1 - \frac{N_{\mathrm{B}}(T_{\mathrm{S,TX}})}{N_{\mathrm{TX}}} \;.
  \label{eq:pswitch}
\end{align}
Note that the impact of other possible chemical reactions, e.g., spontaneous switching, i.e., B $\,\xrightarrow{k_{\mathrm{s}}}\,$ A with rate $k_{\mathrm{s}}$, is neglected in this paper and will be studied in future work.

\scaleSection
\vspace*{+0.15cm}
\section{Analytical End-to End Channel Model}
\label{sec:math}
\vspace*{-0.05cm}
\subsection{Channel Impulse Response}\label{math_sec:cir}
\scaleSubsectionBelow
\vspace*{-0.05cm}
The system introduced in \Section{general_system_model} can be modeled as a 1D channel with infinite extent, i.e., $-\infty < z < \infty$, due to the assumptions of uniform flow and reflective boundaries, and the transparent \ac{RX} model. We derive the probability of molecules to be observed at \RXsimp $\mkern+5mu$ after being switched to state A at position $z_{\mathrm{TX}}$ at time instant $t_{\mathrm{BA}}$ as a function of time. We refer to this probability as $h(t)$. Here, $t_{\mathrm{BA}}$ is within the illumination duration $T_{\mathrm{S,TX}}$ for illumination start time $t=0$, i.e., $t_{\mathrm{BA}} \in [0, 0 + T_{\mathrm{S,TX}}]$.
In the following, we approximate $t_{\mathrm{BA}}$ by $t_{\mathrm{BA}}=0$, which is a valid assumption for small $T_{\mathrm{S,TX}}$, i.e., if the end of the modulation time length is still approximately at $t=0$. The position $z_{\mathrm{TX}}$ where the molecule is switched to state A is random and uniformly distributed in $[z_\mathrm{a,TX}, z_\mathrm{b,TX}]$.

\begin{prop}
The probability $h(t)$ that a molecule in state A, which is uniformly distributed inside the \ac{TX} region, $z_\mathrm{a,TX} \leq z_{\mathrm{TX}} \leq z_\mathrm{b,TX}$, at $t=0$, is observed by \RXsimp $\mkern+5mu$with dimension $z_\mathrm{a,RX} \leq z_{\mathrm{RX}} \leq z_\mathrm{b,RX}$ is given by
\scaleAlign
\vspace*{-0.02cm}
\begin{align}
  &h(t) \mkern-5.5mu=\mkern-5.5mu \frac{1}{2 \,l_{\mathrm{TX}}}\mkern-5.5mu \sum_{i=0}^3 (\mkern-3.5mu-\mkern-3.5mu1)^{i} \mkern-5.5mu\left[\mkern-2.5mu  a_i \erf\mkern-4.5mu\left(\mkern-5.5mu\frac{a_i}{\sqrt{\mkern-3.5mu 4 D_{\mathrm{A}} t}}\mkern-5.5mu\right) \mkern-5.5mu + \mkern-5.5mu \sqrt{\mkern-3.5mu\frac{4 D_{\mathrm{A}} t}{\pi}} \mkern-3.5mu \exp\mkern-5.5mu\left(\mkern-5.5mu \frac{- a_i^2}{4 D_{\mathrm{A}} t}\mkern-5.5mu\right)\mkern-5.5mu   \right] \mkern+5.5mu,\nonumber \\[-0.4cm]
   \label{cir_analytic}
\end{align}
where $\{ a_0, a_1, a_2, a_3\} = \{ z_\mathrm{b,RX} - z_\mathrm{a,TX} - v t, z_\mathrm{b,RX} - z_\mathrm{b,TX} - v t, z_\mathrm{a,RX} - z_\mathrm{b,TX} - v t, z_\mathrm{a,RX} - z_\mathrm{a,TX} - v t \}$. Here, $\erf(x)$ denotes the Gaussian error function.
\end{prop}

\begin{IEEEproof}
  The probability $C_{\mathrm{A}}(t, z_{\mathrm{RX}}, z_{\mathrm{TX}}) = \frac{1}{\sqrt{4 \pi D_{\mathrm{A}} t}} \exp\left( - \frac{(z_{\mathrm{RX}}-z_{\mathrm{TX}} - v t)^2}{4 D_{\mathrm{A}} t}\right)$ that one molecule released at $z_{\mathrm{TX}}$ is observed at $z_{\mathrm{RX}}$ can be derived from \cite[Eq.~(18)]{Jamali2019ChannelMF} by integrating over the \mbox{$x$-$y$} plane.
  The probability of a molecule in state A to be observed at \RXsimp $\mkern+5mu$ is obtained by marginalization over the release position $z_{\mathrm{TX}}$ and integration over the axial extent of \RXsimp $\mkern+5mu$ as follows
  \scaleAlign
    \begin{align}
    &h(t) = \underset{z_{\mathrm{TX}}}{\mathbb{E}} \left\{ \int \limits_{z_\mathrm{a,RX}}^{z_\mathrm{b,RX}} C_{\mathrm{A}}(t, z_{\mathrm{RX}}, z_{\mathrm{TX}}) \dzr \right\} \nonumber \\[-0.1cm]
    &=\int \limits_{z_\mathrm{a,TX}}^{z_\mathrm{b,TX}} \int \limits_{z_\mathrm{a,RX}}^{z_\mathrm{b,RX}} C_{\mathrm{A}}(t, z_{\mathrm{RX}}, z_{\mathrm{TX}}) \fpdf[z_{\mathrm{TX}}]{z_{\mathrm{TX}}} \dzr \dzm \nonumber \\[-0.1cm]
    &\overset{{(a)}}{=}\mkern-6mu \frac{1}{2 \,l_{\mathrm{TX}}}\mkern-11mu\int \limits_{z_\mathrm{a,TX}}^{z_\mathrm{b,TX}} \mkern-15mu \erf\mkern-5.5mu\left(\mkern-5.5mu \frac{z_\mathrm{b,RX}\mkern-5.5mu -\mkern-5.5mu z_{\mathrm{TX}}\mkern-5.5mu - \mkern-5.5muv t}{\sqrt{4 D_{\mathrm{A}} t}}\mkern-4.5mu\right)\mkern-5.5mu -\mkern-4.5mu \erf\mkern-5.5mu\left( \mkern-5.5mu\frac{z_\mathrm{a,RX}\mkern-5.5mu - \mkern-5.5muz_{\mathrm{TX}}\mkern-5.5mu -\mkern-5.5mu v t}{\sqrt{4 D_{\mathrm{A}} t}}\mkern-4.5mu\right)\mkern-4.5mu \dzm \,, \nonumber\\[-0.55cm]
  \end{align}
    where $\mathbb{E}\{\cdot \}$ denotes the expectation operator, and we exploit in $(a)$ the uniform distribution of $z_{\mathrm{TX}}$, i.e., $\fpdf[z_{\mathrm{TX}}]{z_{\mathrm{TX}}} = \frac{1}{l_{\mathrm{TX}}} $, $z_\mathrm{a,TX} \leq z_{\mathrm{TX}} \leq z_\mathrm{b,TX}$, and $\erf(x, y) = \erf(y) - \erf(x) =  \frac{2}{\sqrt{\pi}} \int_x^y \exp\left(-z^2 \right)  \,\mathrm{d}  z$. Finally, we use the integral $\int \erf(x) \dx = x \erf(x) + \frac{1}{\sqrt{\pi}} \exp\left( -x^2\right)$ to obtain \Equation{cir_analytic}.
\end{IEEEproof}
\vspace*{0.08cm}
\scaleSubsection
\subsection{Statistical Model}\label{ssSec:IM}
\scaleSubsectionBelow
In this section, we determine the statistics of the number of signaling molecules, initially in state B, observed as state A molecules at \RXsimp. The end-to-end channel is characterized by a triple stochastic process, where the output of one process is the input of the subsequent process. In each process, a subset of the input molecules is selected, where the actual number follows a Binomial distribution, as the molecules in each process can be modeled as statistically independent.
\scaleSubsubsection
\subsubsection{Number of molecules at \ac{TX}}\label{ssSec:InTX}
\scaleSubsubsectionBelow
$N_{\mathrm{sys}}$ signaling molecules are initially uniformly distributed in subvolume S and in state B. However, only a subset of the signaling molecules $N_{\mathrm{TX}}$, i.e., $N_{\mathrm{TX}} \leq N_{\mathrm{sys}}$, is in $S_{\mathrm{TX}}$ at $t=0$, i.e., with a probability of $ p_{\mathrm{TX}} =  \frac{V_{\mathrm{TX}}}{V_{\mathrm{sys}}}$ a given signaling molecule M is within the \ac{TX} volume, i.e., $\mathrm{M} \in S_{\mathrm{TX}}$, and with a probability of $1-p_{\mathrm{TX}}$ it is outside the \ac{TX}. Hence, the availability of a molecule for information transmission can be modeled as a Bernoulli random variable.
The total number of molecules $N_{\mathrm{TX}}$ available at the \ac{TX} therefore follows the Binomial distribution
\scaleAlign
\begin{align}
  N_{\mathrm{TX}}  \sim \binomial{N_{\mathrm{sys}}, p_{\mathrm{TX}}} \;,\vspace*{-0.5cm}
\end{align}
where $\binomial{N, p}$ denotes a binomial distribution with parameters $N$ and $p$. Here, $N$ and $p$ denote the number of trials and the success probability, respectively.
\scaleSubsubsection
\subsubsection{Number of switched molecules}\label{ssSec:Switched}
\scaleSubsubsectionBelow
 When a state B molecule is within $S_{\mathrm{TX}}$ at $t=0$, it is switched to a state A molecule with probability $s \mkern2mu p_{\mathrm{switch}}$ within the modulation interval. From \Equation{diff_solution}, we observe that $p_{\mathrm{switch}}$ depends on the initial number of molecules $N_{\mathrm{TX}}$ in $S_{\mathrm{TX}}$. In particular, for large $N_{\mathrm{TX}}$, $p_{\mathrm{switch}}$ is dominated by the competition of signaling molecules for photons. Therefore, in the switching process, there exists a dependence between the states of different molecules, cf. \Equation{diff_solution}. On the other hand, for small $N_{\mathrm{TX}}$, the competition can be neglected. In this case, $p_{\mathrm{switch}}$ is constant \ac{w.r.t.} $N_{\mathrm{TX}}$, which is what we will assume in the following. The validity of this assumption is verified in \Section{sec:switching_process_verification}. Hence, the overall number of state A molecules at the \ac{TX} after the modulation follows the Binomial distribution
 \scaleAlign
 \begin{align}
  N_{\mathrm{A}} \mkern-5mu \sim \binomial{\mkern-4mu N_{\mathrm{sys}}, s \mkern2mu p_{\mathrm{TX}} \mkern2mu p_{\mathrm{switch}}\mkern-4mu}\mkern2mu. \vspace*{-0.5cm}
\end{align}
In the following, we refer to the difference between the expectation, $\mathbb{E}\{N_{\mathrm{A}}\} = N_{\mathrm{sys}}  s \mkern2mu p_{\mathrm{TX}} \mkern2mu p_{\mathrm{switch}}$, and the realization, $N_{\mathrm{A}}$, as \ac{TX} noise $n_{\mathrm{TX}}$, i.e., $n_{\mathrm{TX}}= N_{\mathrm{A}} - N_{\mathrm{sys}}  s \mkern2mu p_{\mathrm{TX}} \mkern2mu p_{\mathrm{switch}}$. The transmitter noise is a key characteristic of media modulation based \ac{MC}, where the number of signaling molecules used for information transmission can not be controlled as the molecules are not released. Hence, $n_{\mathrm{TX}}$ is affected by both, the randomness of the availability of the signaling molecules at the \ac{TX} and the randomness of the switching process.

\scaleSubsubsection
\subsubsection{Number of received molecules}\label{ssSec:arrived}
\scaleSubsubsectionBelow
Finally, the number of received state A molecules $N_{\mathrm{RX}}$ at \RXsimp $\mkern+5mu$at time $t_{\mathrm{s}}$ additionally depends on the molecule propagation in the channel. As all molecules are assumed to be independent during propagation, the arrival of a state A molecule at \RXsimp $\mkern+5mu$can be modeled as a Bernoulli random variable with success probability $h(t)$ according to \Equation{cir_analytic}. Hence, $N_{\mathrm{RX}}$ follows the Binomial distribution
\scaleAlign
\vspace*{-0.1cm}
\begin{align}
  N_{\mathrm{RX}}(t = t_{\mathrm{s}})  \sim  \binomial{N_{\mathrm{sys}},p_{\mathrm{r}}}\;,\vspace*{-0.5cm}
  \label{nrMolRX}
\end{align}
where $  p_{\mathrm{r}} = s \mkern2mu p_{\mathrm{TX}} \mkern2mu p_{\mathrm{switch}} h(t = t_{\mathrm{s}})$. Finally, the end-to-end \ac{CIR} is given as $\overline{N}_{\mathrm{RX}}(t) = \mathbb{E}\{N_{\mathrm{RX}}(t) \given s = 1\} = N_{\mathrm{sys}} \mkern2mu p_{\mathrm{TX}} \mkern2mu p_{\mathrm{switch}} h(t)$ as a function of time $t$.

\scaleSection
\vspace*{0.1cm}
\section{Symbol Detection and Performance Analysis}
\vspace*{0.1cm}
\label{sec:performance}
\vspace*{-0.15cm}
In this section, we provide a threshold based detection scheme and derive the \ac{BER} for media modulation based \ac{MC}.
\scaleSubsection
\vspace*{-0.3cm}
\subsection{Detector}\label{math_sec:ssDet}
\scaleSubsectionBelow
\vspace*{-0.01cm}

According to the proposed system model, we assume that initially all signaling molecules are in state B and the state of these molecules can only be switched by illumination at the \ac{TX}. Hence, for transmit bit $s=0$, probability $s \mkern2mu p_{\mathrm{switch}} = 0$ follows, and according to \Equation{nrMolRX} zero molecules in state A are observed. On the other hand, once the receiver counts at least one state A molecule, the receiver should estimate $\hat{s} = 1$. Therefore, we employ the following threshold based decision rule
\scaleAlign
\vspace*{-0.2cm}
  \begin{align}
    \hat{s} &=
        \begin{cases}
          1 , \; \text{if} \; N_{\mathrm{RX}}(t = t_{\mathrm{s}}) \geq  \theta \\[-0.1cm]
          0 , \;\text{otherwise} \;
        \end{cases}\;,
        \label{eq:math_section:threshold}
  \end{align}
  with threshold $\theta = 1$.
\scaleSubsection
\vspace*{0.08cm}
\subsection{Bit Error Rate}
\label{Bit_error_rate_derivation}
\scaleSubsectionBelow
\vspace*{-0.08cm}
According to \Equation{eq:math_section:threshold}, the transmission of bit $s=0$ is error free. The transmission error probability for $s=1$ is non-zero due to the probability, that none of the randomly distributed, randomly switched, and randomly propagating molecules is observed at \RXsimp $\mkern+5mu$ at $t = t_{\mathrm{s}}$.
Therefore, the \ac{BER} can be expressed as
\scaleAlign
\begin{align}
  P_{\mathrm{e}} &= \frac{1}{2} \underbrace{\prob{\hat{s} = 1 \given s = 0}}_{= 0} + \frac{1}{2} \prob{\hat{s} = 0 \given s = 1} \nonumber \\[-0.1cm]
  &= \frac{1}{2} \binom{N_{\mathrm{Sys}}}{0} p_{\mathrm{r}}^0 (1-p_{\mathrm{r}})^{N_{\mathrm{Sys}}} = \frac{1}{2} (1-p_{\mathrm{r}})^{N_{\mathrm{Sys}}} \;. \label{BER_derivation}
\end{align}

\scaleSection
\vspace*{0.1cm}
\section{Performance Evaluation}
\label{sec:evaluation}
\vspace*{-0.1cm}
In this section, we first specify the properties of \ac{GFPD} \cite{brakemann2011reversibly}, which we investigate as a practically feasible option for photoswitchable fluorescent molecules. Then, we evaluate the statistical model in \Equation{nrMolRX} and compare it to results from \ac{PBS}. Finally, the analytical expression for the \ac{BER} in \Equation{BER_derivation} is evaluated for different system configurations.
%
%%%%%%%%%%%%%%%%%%%%%%%%%%%%%%%%%%%%%%%%%%%%%%%%%%%%%%%%%%%%%%%%%%%%%%%%%%%%%%%%
%
\scaleSubsection
\subsection{Choice of Parameter Values}\label{section:evaluation:PBSAndParam}
\scaleSubsectionBelow

The default values of the channel parameters are given in \Table{Table:Parameter} and are used if not specified otherwise. We consider \ac{GFPD} as signaling molecules, as \ac{GFPD} possesses the properties needed according to \Section{sec:model}. The \ac{GFPD} specific parameter values are taken from \cite{brakemann2011reversibly, Junghans2016DiffusionGFPD}. The parameters related to the duct are chosen such that they have the same order of magnitude as those found in a cardiovascular system \cite[Chap. 14]{hall2020guyton}.
To verify the accuracy of the analytical expression for the statistics of the received molecules in \Equation{nrMolRX}, stochastic \ac{PBS} were carried out. The results from \ac{PBS} were averaged over $10^{4}$ realizations.
\begin{table}[!tbp]
\caption{Default Values for Simulation Parameters.}
\vspace*{-0.4cm}
\begin{center}
{\def\arraystretch{1.3}\tabcolsep=2pt
 \begin{tabular}{|l | c | c | r|}
   \hline
 Parameter & Description & Value & Ref. \\ [0.5ex]
 \hline\hline
 $ L_{\mathrm{sys}}$ & Pipe length of S & $0.5 \,\si{\meter}$ &  \\
 \hline
 $H$ & Pipe height & $0.001 \,\si{\meter}$ & \cite{hall2020guyton} \\
 \hline
 $W$ & Pipe width & $0.001 \,\si{\meter}$ & \cite{hall2020guyton}\\
 \hline
 $l_{\mathrm{TX}}$ & \ac{TX} length & $0.05 \,\si{\meter}$ &\\
 \hline
 $d$ & Distance between \ac{TX} and \ac{RX} & $ 0.2 \,\si{\meter}$ & \\
 \hline
 $l_{\mathrm{RX}}$ & \ac{RX} length & $ 0.05\, \si{\meter}$ &\\
 \hline
 $v$ & Flow velocity &  $0.01 \, \si{\meter \per \second}$ & \cite{hall2020guyton}\\
 \hline
 $D_{\mathrm{A}}$ & Diffusion coefficient & $1 \times 10^{-10}\, \si{\meter \squared \per \second} $ &\cite{Junghans2016DiffusionGFPD} \\
 \hline
 $ N_{\mathrm{sys}}$ & Number of molecules in S & $1000$ & \\
 \hline
 $\lambda_{\mathrm{BA}}$ & Wavelength to switch GFP at \ac{TX} & $365 \times 10^{-9} \,\si{\meter}$ &\cite{brakemann2011reversibly}\\
 \hline
 $\lambda_{\mathrm{AB}}$ & Wavelength to switch GFP at \ac{EX} & $405 \times 10^{-9} \,\si{\meter}$ &\cite{brakemann2011reversibly}\\
 \hline
 $\lambda_{\mathrm{f, in}}$ & Wavelength to trigger fluorescence & $515 \times 10^{-9} \,\si{\meter}$ &\cite{brakemann2011reversibly}\\
 \hline
 $\lambda_{\mathrm{f, out}}$ & Wavelength of fluorescence & $529 \times 10^{-9} \,\si{\meter}$ &\cite{brakemann2011reversibly}\\
 \hline
 $\epsilon$ & Molar absorption coefficient & $8.3 \times 10^{3} \,\si{\meter^2 \mathrm{mol}^{-1}}$ &\cite{brakemann2011reversibly}\\
 \hline
 $\varphi_{\mathrm{B}}$ & Quantum yield &  $0.41$ &\cite{brakemann2011reversibly}\\
 \hline
 $T_{\mathrm{S,TX}}$ & Irradiation time at \ac{TX} & $ 5 \times 10^{-3} \,\si{\second}$ &\cite{brakemann2011reversibly}\\
 \hline
 $\Delta t$ & Time step \ac{PBS} & $ 1 \times 10^{-2} \,\si{\second}$ &\\
 \hline
 \end{tabular}
 }
\end{center}
 \label{Table:Parameter}
 \vspace*{-0.6cm}
\end{table}

%%%%%%%%%%%%%%%%%%%%%%%%%%%%%%%%%%%%%%%%%%%%%%%%%%%%%%%%%%%%%%%%%%%%%%%%%%%%%%%%
\vspace*{+0.1cm}
\scaleSubsection
\subsection{Evaluation of the Switching Process} \label{sec:switching_process_verification}
\scaleSubsectionBelow
\begin{figure}[!tbp]
  \centering
  \includegraphics[width = 0.93\columnwidth, trim={0 0 0 1.3cm},clip]{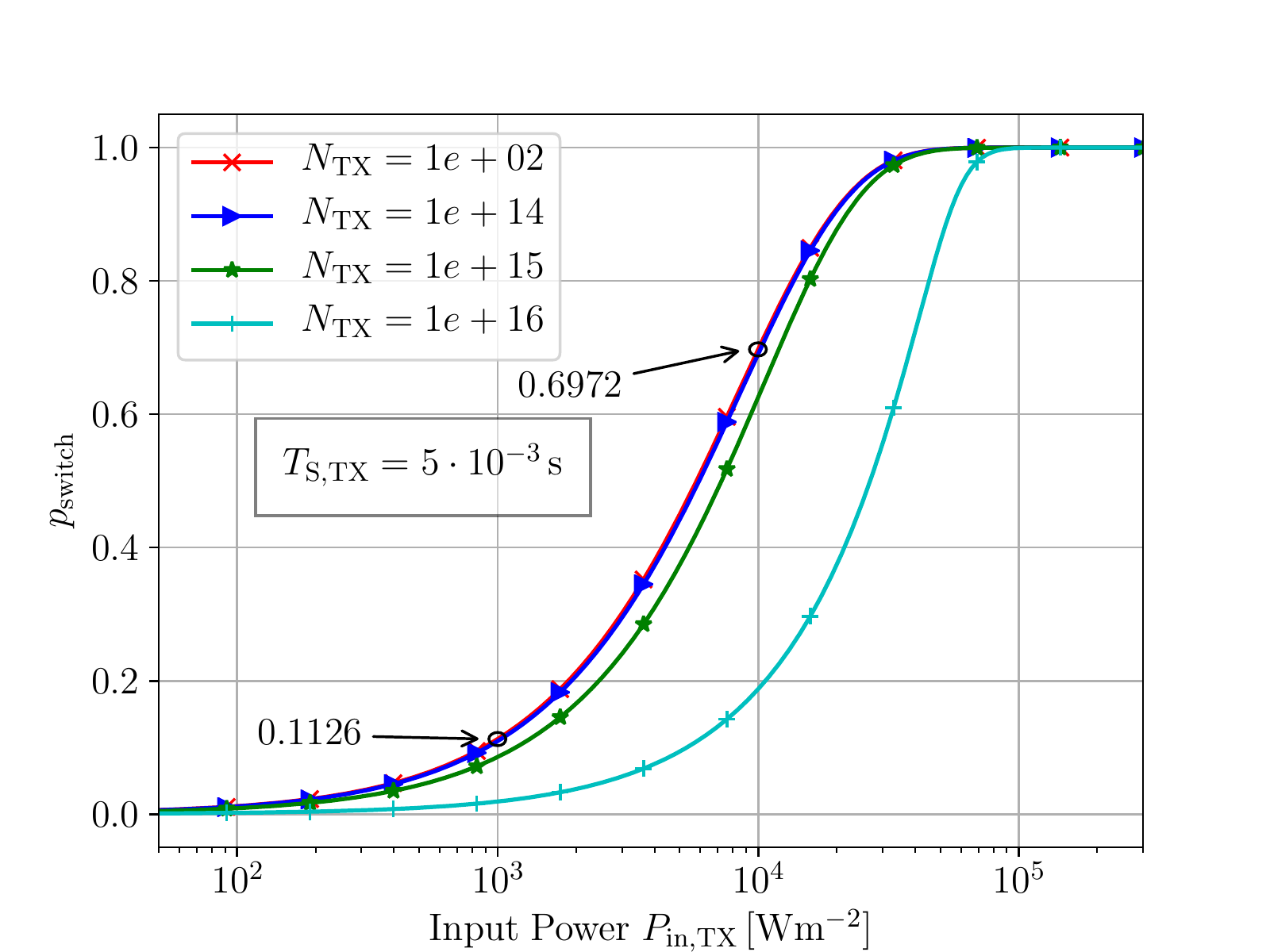}\vspace*{-0.5mm}
  \caption{The probability of a molecule within the \ac{TX} to switch from state B to state A within the modulation time length $T_{\mathrm{S,TX}}$ as a function of the input power $P_{\mathrm{in, TX}}$ for different numbers of molecules $N_{\mathrm{TX}}$ in $S_{\mathrm{TX}}$.}
  \label{fig:switching_characteristic}
  \vspace*{-1mm}
\end{figure}
First, we investigate the photochemical reaction, discussed in \Section{photoChemic}, for \ac{GFPD}. In \Figure{fig:switching_characteristic}, $p_{\mathrm{switch}}$, as defined in \Equation{eq:pswitch}, is shown as a function of input light power $P_{\mathrm{in, TX}}$ with irradiation duration $ T_{\mathrm{S,TX}} = 5 \times 10^{-3} \,\si{\second}$.  $ T_{\mathrm{S,TX}}$ is chosen such that it is in the range of values considered in \cite{brakemann2011reversibly}, where an irradiation duration between $1 \times 10^{-7} \,\si{\second}$ and $5 \,\si{\second}$ was used. Note that condition \Equation{eq:assumption_propagation_length} is satisfied, since $ 5.1 \times 10^{-5} \,\si{\meter} \ll  5 \times 10^{-2} \,\si{\meter} $. Moreover, the range of values for $P_{\mathrm{in, TX}}$ considered here is also reasonable according to \cite{brakemann2011reversibly}, where power values ranging from $ 1 \times 10^{3} \,\si{\watt \per\m \squared}$ to  $ 1.6 \times 10^{6} \, \si{\watt \per\m \squared}$ were used for light sources with wavelength $\lambda_{\mathrm{BA}}$.
% In addition, the authors in \cite{brakemann2011reversibly} show the feasibility of lasers as light sources, e.g. for imaging, with light power values up to $ 4.3 \times 10^{10} \, \si{\watt \per\m \squared}$.
From \Figure{fig:switching_characteristic}, we observe that the likelihood of a molecule to switch within the irradiation time is low for small input power, increases for increasing input power, and converges to $1$ for large values of $P_{\mathrm{in, TX}}$.
Moreover, \Figure{fig:switching_characteristic} shows that $p_{\mathrm{switch}}$ remains unchanged for a large range of $N_{\mathrm{TX}}$. Only for systems with a very large number of signaling molecules, i.e., if $N_{\mathrm{TX}}\geq 10^{14}$, a larger input power is necessary to achieve a given switching probability due to the competition for photons among the signaling molecules. Thus, approximating $p_{\mathrm{switch}}$ to be independent of $N_{\mathrm{TX}}$ is possible for the system under investigation if $N_{\mathrm{TX}}$ is sufficiently small, see \Section{ssSec:Switched}. Therefore, as $N_{\mathrm{TX}}$ is random and upper bounded by $N_{\mathrm{sys}}$, $N_{\mathrm{TX}}\leq N_{\mathrm{sys}} \leq 10^{14}$ is sufficient to ensure statistical independence of the signaling molecules.
\vspace*{+0.1cm}
\scaleSubsection
\subsection{Evaluation of the Statistics of $N_{\mathrm{RX}}$} \label{sec:statistic_verification}
\scaleSubsectionBelow
\begin{figure}[!tbp]
  %\vspace*{-1mm}
  \hspace{0.3cm}\includegraphics[width = 0.93\columnwidth, trim={0 0 0 1.3cm},clip]{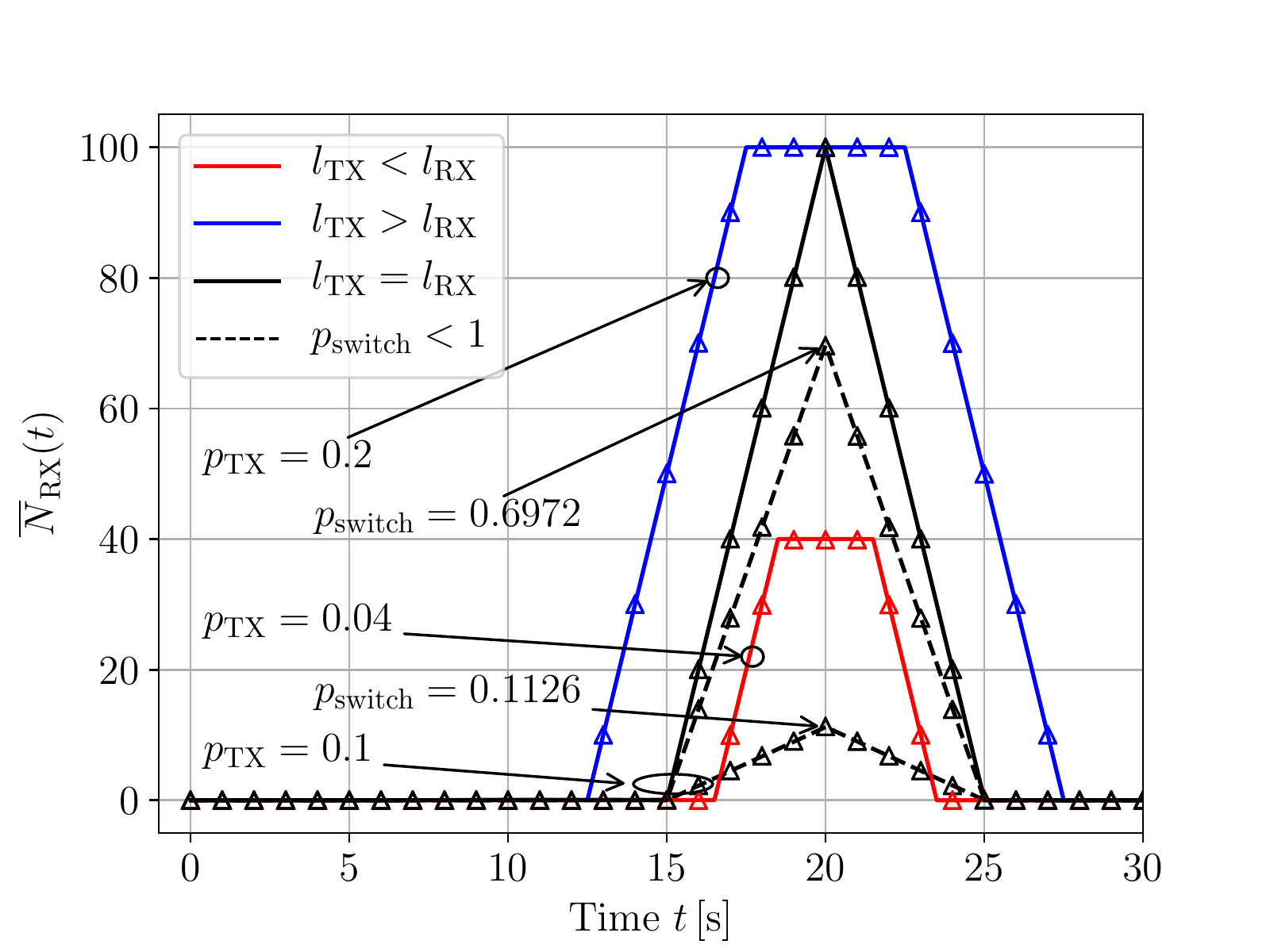}\vspace*{-0.5mm}
  \caption{End-to-end channel impulse response $\overline{N}_{\mathrm{RX}}(t)$ for $l_{\mathrm{TX}} = l_{\mathrm{RX}}$ (black), $ l_{\mathrm{TX}} \neq l_{\mathrm{RX}}$ (blue and red) with $ l_{\mathrm{TX}} = 0.1 \, \si{\meter}$ and $ l_{\mathrm{TX}} = 0.02 \, \si{\meter}$, respectively, and switching probabilities $p_{\mathrm{switch}}$ obtained for $P_{\mathrm{in, TX}} = 10^{3} \,\si{\watt \per\m \squared}$, $P_{\mathrm{in, TX}} = 10^{4} \,\si{\watt \per\m \squared}$ (both dashed), and $P_{\mathrm{in, TX}} = 10^{5} \,\si{\watt \per\m \squared}$ (solid), see \Figure{fig:switching_characteristic}. The results obtained from \ac{PBS} are depicted by triangle markers.}
  \label{fig:mean_statistic}
  \vspace*{-0.3cm}
\end{figure}
\begin{figure}[!tbp]
  \includegraphics[width = 0.93\columnwidth, trim={0 0 0 1.3cm},clip]{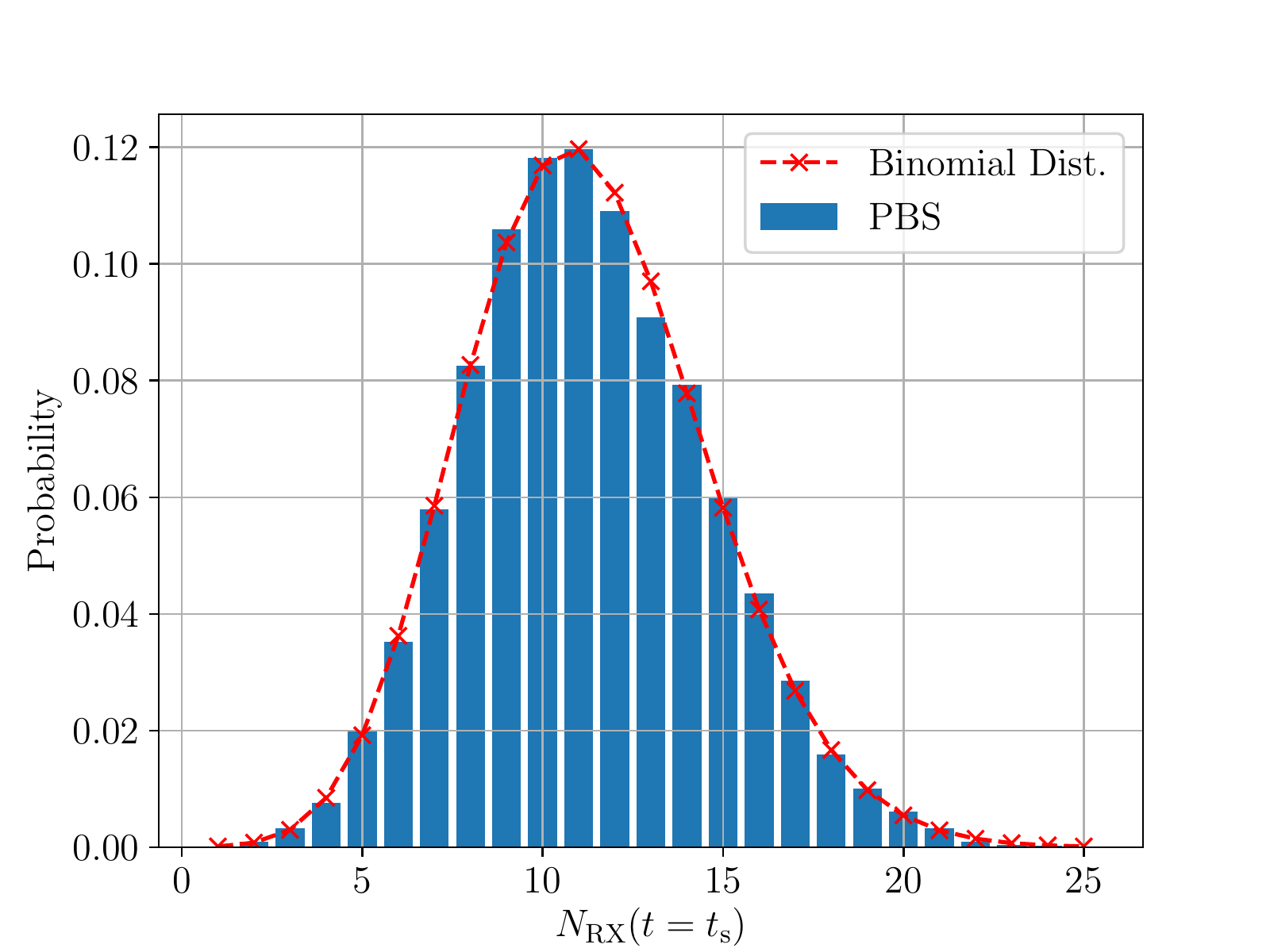}\vspace*{-0.5mm}
  \caption{The empirical (PBS) and analytical distribution of the number of received molecules $N_{\mathrm{RX}}(t = t_{\mathrm{s}})$ according to \Equation{nrMolRX} for $s=1$ and $p_{\mathrm{switch}} = 0.1126$.}
  \label{fig:histogram}
  \vspace*{-0.3cm}
\end{figure}
In \Figure{fig:mean_statistic}, we show the end-to-end \ac{CIR} $\overline{N}_{\mathrm{RX}}(t)$ derived in \Section{ssSec:arrived}, and compare it to results obtained from \ac{PBS}. We observe that the analytical results are in excellent agreement with the \ac{PBS} results. We further observe that for $ l_{\mathrm{TX}} = l_{\mathrm{RX}}$, $\overline{N}_{\mathrm{RX}}(t)$ has a unique peak at sampling time $t = t_{\mathrm{s}} =  20 \, \si{\second}$ and the peak height scales linearly with $p_{\mathrm{switch}}$. Moreover, for larger \ac{TX} lengths, i.e, $l_{\mathrm{TX}} > l_{\mathrm{RX}}$ (blue), the extent of the modulated molecules is larger than the axial extent of the transparent \ac{RX}. Therefore, not all modulated molecules can be detected by the \RXsimp  $\mkern+5mu$ at once, and hence, $\overline{N}_{\mathrm{RX}}(t)$ is constant around the sampling time.
In contrast, for smaller \ac{TX} lengths, i.e., $l_{\mathrm{TX}} < l_{\mathrm{RX}}$ (red), $\overline{N}_{\mathrm{RX}}(t_{\mathrm{s}})$ decreases, which is intuitive as, for a smaller \ac{TX}, $p_{\mathrm{TX}}$ is smaller, i.e., fewer signaling molecules $N_{\mathrm{TX}}$ are within $S_{\mathrm{TX}}$.
In \Figure{fig:histogram}, we show the probability mass function of the number of received molecules $ N_{\mathrm{RX}}(t_{\mathrm{s}})$ for $s=1$ according to \Equation{nrMolRX} and compare it to results obtained by \ac{PBS}. We observe that the results obtained from \ac{PBS} and \Equation{nrMolRX}, respectively, match, which confirms the statistical model proposed in \Section{ssSec:IM}.

\vspace*{+0.1cm}
\scaleSubsection
\subsection{Evaluation of BER} \label{sec:BER_plots}
\scaleSubsectionBelow
\begin{figure}[!tbp]
  \includegraphics[width = 0.93\columnwidth, trim={0 0 0 1.3cm},clip]{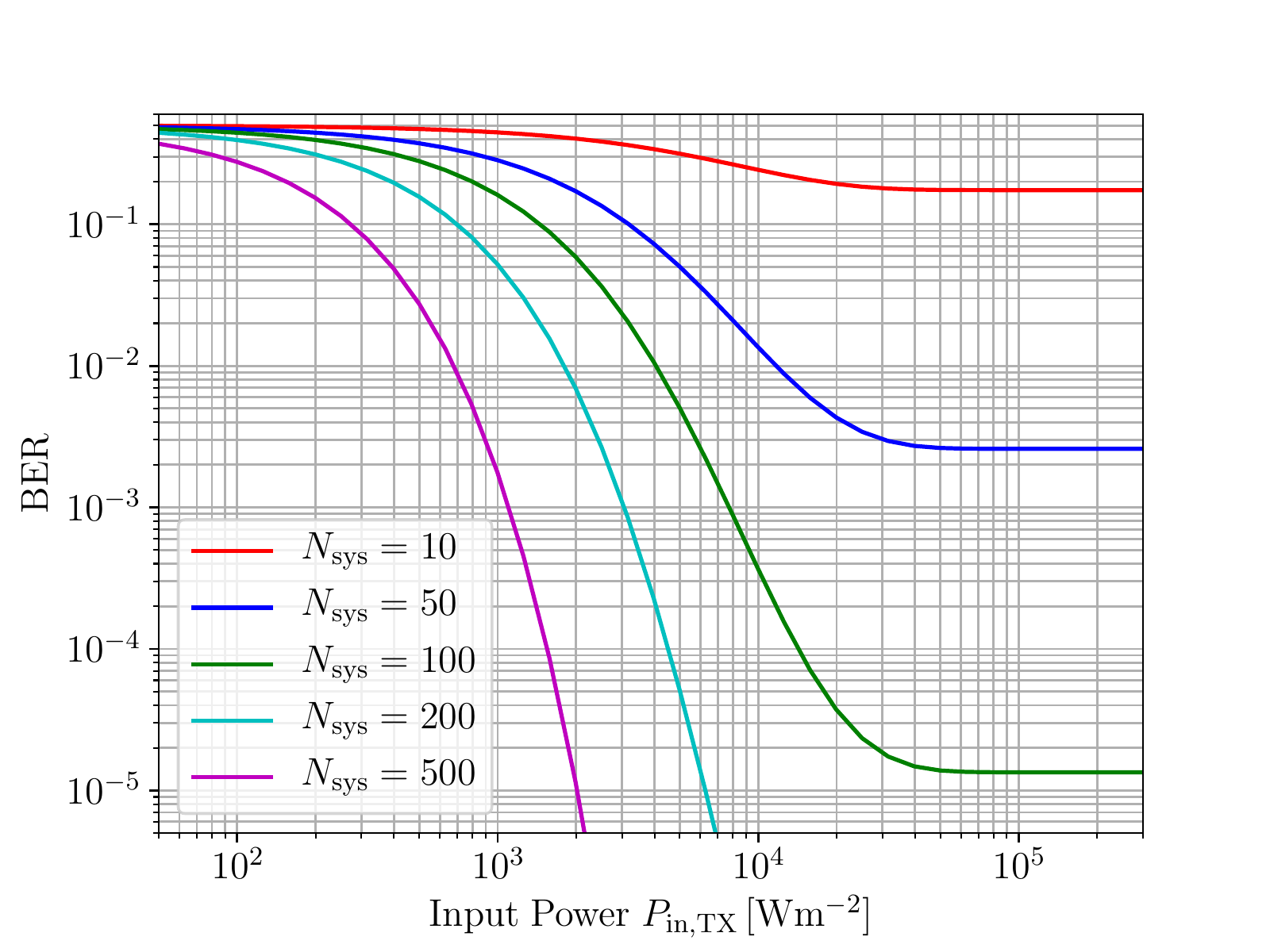}\vspace*{-0.5mm}
  \caption{\ac{BER} as a function of the irradiation power for different numbers of signaling molecules.}
  \label{fig:BER_over_power}
    \vspace*{-0.4cm}
\end{figure}
In \Figure{fig:BER_over_power}, the \ac{BER} is shown as a function of the irradiation power $P_{\mathrm{in, TX}}$ for different numbers of signaling molecules $N_{\mathrm{sys}}$. Here, $h(t = t_{\mathrm{s}}) = 0.999 $ and $p_{\mathrm{TX}} = 0.1$.
We observe that \ac{BER} decreases as $P_{\mathrm{in, TX}}$ increases. For $P_{\mathrm{in, TX}} > 6 \times 10^{4} \, \si{\watt \per\m \squared}$, the \ac{BER} approaches an error floor, which is visible for $N_{\mathrm{sys}} = \{10, 50, 100\}$, but also occurs for larger $N_{\mathrm{sys}}$ at lower \ac{BER} values. In particular, for large power values, $p_{\mathrm{switch}} = 1$ follows, i.e., the switching is deterministic. However, the \ac{BER} exhibits an error floor due to the \ac{TX} noise $n_{\mathrm{TX}}$ caused by the randomness of the actual number of signaling molecules, $N_{\mathrm{TX}}$, available at the \ac{TX}, cf. \Section{ssSec:Switched}. Moreover, we note that in agreement with \Equation{BER_derivation}, the \ac{BER} decreases as $N_{\mathrm{sys}}$ increases.

\scaleSection
\vspace*{0.1cm}
\section{Conclusion}
\label{sec:conclusion}
\vspace*{-0.2cm}
In this paper, we introduced a new form of media modulation for \ac{MC}. Media modulation does not require a \ac{TX} that stores signaling molecules and controls their release. In particular, in media modulation based \ac{MC}, a \ac{TX} is utilized which alters the state of signaling molecules already present in the channel. We investigated the properties of media modulation for the group of photochromic molecules, whose states can be controlled by external light stimuli. Furthermore, we studied the usage of these molecules for information transmission in a \ac{3D} duct system with one \ac{TX} and one \ac{RX}. We developed a statistical model for the received signal taking into account the randomness of the initial molecule distribution, i.e., their availability at the \ac{TX}, the randomness of the switching process, and the randomness of molecule propagation. Finally, we analyzed the performance of a transmission link based on media modulation in terms of \ac{BER}. Our numerical results showed that media modulation enables reliable information transmission. Therefore, media modulation provides a new perspective for designing non-invasive MC systems.

In this paper, we focused on the \ac{TX} modelling for media modulation, while a detailed investigation of the sensing process at the RX and the back-switching of signaling molecules by an \ac{EX} are interesting topics left for future work. Furthermore, it would be interesting to analyze the applicability of other biological processes, such as phosphorylation, for synthetic media modulation based \ac{MC}.

\vspace*{-0.1cm}
\bibliographystyle{IEEEtran}
\bibliography{literature}

\begin{thebibliography}{10}
\providecommand{\url}[1]{#1}
\csname url@samestyle\endcsname
\providecommand{\newblock}{\relax}
\providecommand{\bibinfo}[2]{#2}
\providecommand{\BIBentrySTDinterwordspacing}{\spaceskip=0pt\relax}
\providecommand{\BIBentryALTinterwordstretchfactor}{4}
\providecommand{\BIBentryALTinterwordspacing}{\spaceskip=\fontdimen2\font plus
\BIBentryALTinterwordstretchfactor\fontdimen3\font minus
  \fontdimen4\font\relax}
\providecommand{\BIBforeignlanguage}[2]{{%
\expandafter\ifx\csname l@#1\endcsname\relax
\typeout{** WARNING: IEEEtran.bst: No hyphenation pattern has been}%
\typeout{** loaded for the language `#1'. Using the pattern for}%
\typeout{** the default language instead.}%
\else
\language=\csname l@#1\endcsname
\fi
#2}}
\providecommand{\BIBdecl}{\relax}
\BIBdecl

\bibitem{Jamali2019ChannelMF}
V.~{Jamali}, A.~{Ahmadzadeh}, W.~{Wicke}, A.~{Noel}, and R.~{Schober},
  ``Channel modeling for diffusive molecular communication—{A} tutorial
  review,'' \emph{Proc. IEEE}, vol. 107, no.~7, pp. 1256--1301, Jul. 2019.

\bibitem{Kuran2011}
M.~S. Kuran, H.~B. Yilmaz, T.~Tugcu, and I.~F. Akyildiz, ``Modulation
  techniques for communication via diffusion in nanonetworks,'' in \emph{IEEE
  Int. Conf. Commun. (ICC)}, Kyoto, Japan, Jun. 2011, pp. 1--5.

\bibitem{Tang2021}
Y.~Tang \emph{et~al.}, ``Molecular type permutation shift keying for molecular
  communication,'' \emph{IEEE Trans. Mol. Biol. Multi-Scale Commun.}, vol.~6,
  no.~2, pp. 160--164, Nov. 2020.

\bibitem{Khandani2013}
A.~K. Khandani, ``Media-based modulation: A new approach to wireless
  transmission,'' in \emph{2013 IEEE Int. Symp. Inf. Theory}, Jul. 2013, pp.
  3050--3054.

\bibitem{Basar2019}
E.~Basar, ``Media-based modulation for future wireless systems: A tutorial,''
  \emph{IEEE Wireless Commun.}, vol.~26, no.~5, pp. 160--166, Oct. 2019.

\bibitem{gohari2016information}
A.~Gohari, M.~Mirmohseni, and M.~Nasiri-Kenari, ``Information theory of
  molecular communication: Directions and challenges,'' \emph{IEEE Trans. Mol.
  Biol. Multi-Scale Commun.}, vol.~2, no.~2, pp. 120--142, Dec. 2016.

\bibitem{farahnak2020molecular}
M.~Farahnak-Ghazani, M.~Mirmohseni, and M.~Nasiri-Kenari, ``On molecular flow
  velocity meters,'' \emph{Accepted for Publication in IEEE Trans. Mol. Biol.
  Multi-Scale Commun.}, Dec. 2020.

\bibitem{Kim2019}
E.~Kim \emph{et~al.}, ``Redox is a global biodevice information processing
  modality,'' \emph{Proc. IEEE}, vol. 107, no.~7, pp. 1402--1424, Apr. 2019.

\bibitem{grusch2014spatio}
M.~Grusch \emph{et~al.}, ``Spatio-temporally precise activation of engineered
  receptor tyrosine kinases by light,'' \emph{The EMBO Journal}, vol.~33,
  no.~15, pp. 1713--1726, Jul. 2014.

\bibitem{chang2014light}
K.-Y. Chang \emph{et~al.}, ``Light-inducible receptor tyrosine kinases that
  regulate neurotrophin signalling,'' \emph{Nature Communications}, vol.~5,
  no.~1, pp. 1--10, Jun. 2014.

\bibitem{balzani2014photochemistry}
V.~Balzani, P.~Ceroni, and A.~Juris, \emph{Photochemistry and Photophysics:
  Concepts, Research, Applications}.\hskip 1em plus 0.5em minus 0.4em\relax
  Weinheim: John Wiley \& Sons, 2014.

\bibitem{brakemann2011reversibly}
T.~Brakemann \emph{et~al.}, ``A reversibly photoswitchable {GFP}-like protein
  with fluorescence excitation decoupled from switching,'' \emph{Nature
  Biotechnology}, vol.~29, no.~10, pp. 942--947, Sep. 2011.

\bibitem{Junghans2016DiffusionGFPD}
C.~Junghans, F.-J. Schmitt, V.~Vukojević, and T.~Friedrich, ``Diffusion
  behavior of the fluorescent proteins {eGFP} and dreiklang in solvents of
  different viscosity monitored by fluorescence correlation spectroscopy,''
  \emph{Optofluidics, Microfluidics and Nanofluidics}, vol.~3, Jan. 2016.

\bibitem{hall2020guyton}
J.~E. Hall and M.~E. Hall, \emph{Guyton and Hall Textbook of Medical Physiology
  e-Book}.\hskip 1em plus 0.5em minus 0.4em\relax Elsevier Health Sciences,
  2020, vol. 12th ed.

\end{thebibliography}
\end{document}